\documentclass[fleqn,usenatbib]{mnras}

\usepackage{newtxtext,newtxmath}

\usepackage[T1]{fontenc}

\DeclareRobustCommand{\VAN}[3]{#2}
\let\VANthebibliography\thebibliography
\def\thebibliography{\DeclareRobustCommand{\VAN}[3]{##3}\VANthebibliography}

\usepackage{graphicx}	
\usepackage{amsmath}	
\usepackage{amssymb}	

\title[Stellar heating feedback]{Implementation of stellar heating feedback in simulations of star cluster formation: effects on the initial mass function}

\author[Sajay Sunny Mathew, Christoph Federrath]{
Sajay Sunny Mathew,$^{1,2}$\thanks{E-mail: \href{mailto:sajay.18ms0092@ap.iitism.ac.in}{sajay.18ms0092@ap.iitism.ac.in}}
Christoph Federrath$^{1}$\thanks{E-mail: \href{mailto:christoph.federrath@anu.edu.au}{christoph.federrath@anu.edu.au}}
\\
$^{1}$Research School of Astronomy and Astrophysics, Australian National University, Canberra, ACT~2611, Australia\\
$^{2}$Indian Institute of Technology (ISM) Dhanbad, Jharkhand-826004, India\\
}

\date{Accepted XXX. Received YYY; in original form ZZZ}

\pubyear{2020}

\begin{document}

\label{firstpage}
\pagerange{\pageref{firstpage}--\pageref{lastpage}}
\maketitle

\begin{abstract}
Explaining the initial mass function (IMF) of stars is a long-standing problem in astrophysics. The number of complex mechanisms involved in the process of star cluster formation, such as turbulence, magnetic fields and stellar feedback, make understanding and modeling the IMF a challenging task. In this paper, we aim to assert the importance of stellar heating feedback in the star cluster formation process and its effect on the shape of the IMF. We use an analytical sub-grid model to implement the radiative feedback in fully three-dimensional magnetohydrodynamical (MHD) simulations of star cluster formation, with the ultimate objective of obtaining numerical convergence on the IMF. We compare a set of MHD adaptive-mesh-refinement (AMR) simulations with three different implementations of the heating of the gas: 1) a polytropic equation of state (EOS), 2) a spherically symmetric stellar heating feedback, and 3) our newly developed polar heating model that takes into account the geometry of the accretion disc and the resulting shielding of stellar radiation by dust. For each of the three heating models, we analyse the distribution of stellar masses formed in ten molecular cloud simulations with different realizations of the turbulence to obtain a statistically representative IMF. We conclude that stellar heating feedback has a profound influence on the number of stars formed and plays a crucial role in controlling the IMF. We find that the simulations with the polar heating model achieve the best convergence on the observed IMF.
\end{abstract}

\begin{keywords}
ISM: clouds -- ISM: kinematics and dynamics -- magnetohydrodynamics (MHD) -- stars: formation
\end{keywords}



\section{Introduction}

The initial mass function (IMF) is the distribution of stellar masses in a young star cluster. It can be thought of as a probability distribution for the mass of stars when they are born. Observations suggest that the IMF is relatively universal \citep{2018PASA...35...39H}. This has far-reaching implications since different characteristics of a star, like the luminosity and lifetime, are dependent on its initial mass. The idea of a probability distribution for the mass of a star was first put forward by \citet{1955ApJ...121..161S}, who defined the number of stars $N(M)$ as a power-law function of stellar mass $M (\mathrm{M_\odot})$, given by $dN \propto M^{-1.35}\, d\mathrm{log}M\, (M > 1\, \mathrm{M_\odot})$. A recent study by \citet{2005ASSL..327...41C} suggests a log-normal form for sub-solar masses, i.e., for masses less than $1\, \mathrm{M_\odot}$ \citep{1979ApJS...41..513M,1986FCPh...11....1S}, and a Salpeter-like slope for the higher-mass stars. Another prevalent proposal is to represent the IMF as a series of power-laws \citep{2001MNRAS.322..231K}. The IMF has a characteristic or peak mass between $0.2-0.3\, \mathrm{M_\odot}$ \citep{2003PASP..115..763C,2008ApJ...681..365E,2014prpl.conf...53O}, with the brown dwarf cutoff at $0.08\, \mathrm{M_\odot}$. A complete understanding of the IMF requires accurate modeling of the formation of a group of stars. The formation of a star cluster is a vigorous and chaotic process that begins with the gravitational collapse of a molecular cloud. Turbulence, magnetic fields, gravity and stellar feedback play decisive roles in the evolution and morphology of molecular clouds and are therefore crucial ingredients for star formation \citep{2018PhT....71f..38F,2019FrASS...6....7K}.

There have been many analytic and numerical studies in the past decade that highlight the influence of one mechanism over the other in shaping the distribution of stellar masses. \citet{2001MNRAS.324..573B} hold competitive accretion between the stars as the principal mechanism responsible for the observed distribution of stellar masses. \citet{2002ApJ...576..870P} argue that the structural evolution of molecular clouds and the stellar IMF can be explained by turbulent fragmentation, i.e., the process of formation of filaments \citep{2014prpl.conf...27A} and dense cores via supersonic turbulence \citep{Haugb_lle_2018}. They point out that the observed slope for stellar masses above $1\, \mathrm{M_\odot}$ can be identified as a direct consequence of the power-law nature of the velocity power spectrum of supersonic MHD turbulence. Although both the competitive accretion and turbulence-triggered theories present evidences for the possibility of a universal IMF, they fail to explain some of the fundamental physical observations or are based on assumptions that require further analysis. The arguments in the competitive accretion model would mean an accretion period that is too long and is unsuccessful in correctly reproducing the Salpeter slope, while the turbulent fragmentation model depends on the correlation between the core mass function (CMF) and the initial mass function, which is still an open question \citep{Smith_2008,Smith_2009,2011IAUS..270..159H}. Recently, the effects of magnetic fields and protostellar outflows on the IMF are being studied extensively \citep{2014MNRAS.439.3420M,2018MNRAS.476..771C,2019FrASS...6....7K}. Another major candidate that can influence the mass spectrum of stars is the radiative feedback by stars. \citet{2009MNRAS.392.1363B} and \citet{2011ApJ...740...74K} propose that stellar radiative feedback may be responsible for setting a universal characteristic mass of the IMF (see also the recent studies by \citealt{2017JPhCS.837a2007F} and \citealt{2018AAS...23111403G}). However, numerical simulations that study the impact of radiation feedback on the IMF are limited due to the computational cost of radiative transfer simulations. An additional problem is that virtually all simulations that include radiative transfer always solve the radiation equations assuming multiple simplifying approximations \citep{2019FrASS...6...51T}.

Obtaining a numerically converged IMF is an important step towards understanding the observed IMF, and this involves running simulations of star cluster formation, incorporating all the physical mechanisms involved in the process. Recent numerical studies of the IMF remain incomplete as too few stars formed in the simulations to obtain a statistically relevant sample. Moreover, the simulation resolution required to include all the mechanisms is arduous to achieve in studies of the IMF. One way of overcoming these limitations is through the use of sub-resolution or sub-grid scale models that reproduce the effects of different mechanisms, such as stellar feedback. There exist many works in the literature that make use of sub-grid models to study star cluster formation \citep[e.g.,][]{2014ApJ...790..128F,2014MNRAS.439.3420M,2017JPhCS.837a2007F}, with the main goal of enabling large parameter studies of what the IMF might depend on, such as the power spectrum and driving of the turbulence \citep{2016MNRAS.462.4171B,2017MNRAS.465..105L}, the virial parameter, the magnetic field strength, etc \citep{2018A&A...611A..88L,2019A&A...622A.125L}.

Here we present a simple sub-grid model \citep[following-up on the previous work by][]{2017JPhCS.837a2007F} to incorporate direct heating feedback from stars in MHD simulations of star cluster formation. This model takes into account the shielding of the stellar radiative flux by the dust particles in the accretion disc around each protostar. Our current modeling capabilities for magnetic fields, turbulence and gravity, along with the sub-grid models, are expected to lead to a better convergence on the IMF and also enable parameter studies. For this purpose, we develop a fast numerical algorithm to incorporate the main effects of stellar heating feedback without the need for full radiation transport. 

In Section~\ref{sec:method} we explain the simulation methods and setups. Section~\ref{sec:model} introduces our newly developed stellar heating model. Section~\ref{sec:parameters} describes the initial conditions and simulation parameters. In Section~\ref{sec:results}, we compare three models of the heating of the gas: 1) polytropic, 2) spherically-symmetric radiative heating, and 3) our new polar heating model. We investigate the column density and temperature structures, evolution of dynamical quantities and the IMF of the stars formed in 10 simulations for each of the three heating models. Limitations are discussed in Section~\ref{sec:discussions}. Section~\ref{sec:conclude} presents our conclusions and summarises the main results.

\section{Methodology}
\label{sec:method}
\subsection{Magnetohydrodynamical equations}
The numerical modeling is performed by solving the magnetohydrodynamical (MHD) equations including gravity using adaptive mesh refinement (AMR) \citep{1989JCoPh..82...64B} in the FLASH code \citep{2000ApJS..131..273F,2008ASPC..385..145D},
\begin{equation}
\frac{\partial\rho}{\partial t}+\nabla\cdot (\rho \mathbf v) = 0,
\end{equation}
\begin{equation}
\rho\, (\frac{\partial}{\partial t} + \mathbf v \cdot \nabla )\, \mathbf v = \frac{(\mathbf B \cdot \nabla) \mathbf B}{4 \pi} - \nabla P_{\mathrm{tot}} + \rho (\mathbf g + \mathrm{\mathbf{F}_{stir}}), \label{eq:mhd2}
\end{equation}
\begin{equation} 
\frac{\partial \mathbf B} {\partial t} = \nabla \times (\mathbf v \times \mathbf B),   \hspace{4mm}\nabla \cdot \mathbf B = 0,
\end{equation}
where $\rho,\mathbf v, P_{\mathrm{tot}} = P + 1/(8\pi) |\mathbf B |^2, \mathbf B$ and $\mathrm{\mathbf{F}_{stir}}$ denote the gas density, velocity, pressure (aggregate of thermal and magnetic), magnetic field and turbulent-acceleration field, respectively. Here $\mathbf{g}$ is the gravitational acceleration and is the sum of the self-gravity of the gas and the acceleration due to the presence of sink particles (see \S\ref{sec:sink}). 

Radiation-hydrodynamic simulations involve the equation of energy conservation, which contains terms that describe the interaction between the gas and radiation. To account for the radiation field, the radiative transfer equation has to be solved for every time step, which is computationally expensive. However, we close the system by using a polytropic equation of state, i.e., an equation that provides $P$ directly from $\rho$ (see \S\ref{sec:polytropic}). Such an equation approximates previous radiation-hydrodynamic simulations of the collapse of cloud cores to form stars \citep{2000ApJ...531..350M}. To actualize the change in the thermal pressure due to stellar heating, we simply add a space-dependent pressure component to the pressure calculated from the polytropic equation of state (explained in more detail in \S\ref{sec:model_temp}).

\subsection{Turbulence}
Turbulence in molecular clouds is a crucial factor that influences the star formation rate and the star formation efficiency \citep{2012ApJ...761..156F,2013ApJ...763...51F,2014prpl.conf...77P,2019FrASS...6....7K}. On the large scale, supersonic turbulent flows support the clouds against a global collapse, but also generates local compressions or shocked regions, promoting star formation \citep{2004RvMP...76..125M,2007ARA&A..45..565M}.  Supernova explosions and other stellar feedback mechanisms as well as dynamical mechanisms (such as galactic rotation and shear, and accretion) drive compressive modes of turbulence in molecular clouds \citep{2017IAUS..322..123F}. In all our simulations, we include a turbulence driving module that mimics the observed turbulence in real molecular clouds, i.e., driving on the largest scales, producing a velocity power spectrum $\sim k^{-2}$ or equivalently a velocity dispersion -- size relation of $\sigma_v \propto \ell^{1/2}$ \citep{1981MNRAS.194..809L,2002A&A...390..307O,2004ApJ...615L..45H,2011ApJ...740..120R}, consistent with supersonic, compressible turbulence \citep{2010A&A...512A..81F,2013MNRAS.436.1245F}. To establish an acceleration field $\mathrm{\mathbf{F}_{stir}}$, our turbulence driving module imposes a stochastic Ornstein-Uhlenbeck process \citep{1988CF.....16..257E,schmidt}. $\mathrm{\mathbf{F}_{stir}}$ acts as a momentum and energy source term in the MHD equations. We use a mixed driving of turbulence with a turbulence driving parameter $\zeta=0.5$ \citep{2010A&A...512A..81F}, common for clouds in the Milky Way disc \citep{2016ApJ...832..143F}. A mixed driving of turbulence refers to a combination of compressive ($\nabla \times \mathbf{F_{\mathrm{stir}}} = 0,\, \zeta \sim 0)$ and solenoidal ($\nabla \cdot \mathbf{F_{\mathrm{stir}}} = 0,\, \zeta \sim 1)$ modes of driving \citep{2008ApJ...688L..79F,2010HiA....15..404F,2012MNRAS.423.2680M,2015MNRAS.451.1380N}. We note that a parameter study should be performed to determine the role of the turbulence driving mode on the IMF. However, this is out of the scope of the present paper and deserves a dedicated investigation. Here we focus on developing a radiation feedback module for use in such a follow-up study.

\subsection{Sink particles and adaptive mesh refinement}
\label{sec:sink}

Sink particles are used in simulations to model the collapse of dense cores, protostar formation, and subsequent accretion \citep{2010ApJ...713..269F}. It is a sub-resolution model for all the internal properties of an unresolved core + disc + protostar system. A sink particle forms where a computational cell exceeds a pre-defined density threshold and all the gas within a control volume (with the size a Jeans volume at that density) defined around the cell is gravitationally bound and converging towards it. A series of other checks are also performed to avoid the spurious formation of sink particles \citep{2010ApJ...713..269F}. The density threshold or the Jeans resolution density is given by
\begin{equation}
    \rho_{\mathrm{sink}} = \frac{\pi\, c_s^2}{G\, \mathrm{\lambda_J^2}} = \frac{\pi\, c_s^2}{4\,G\, r_{\mathrm{sink}}^2},
\end{equation}
where $c_s^2$ is the sound speed, $G$ is the gravitational constant, $\mathrm{\lambda_J}=[\pi c_s^2/(G\rho)]^{1/2}$ is the local Jeans length, and $r_{\mathrm{sink}}= \lambda_\mathrm{J}/2$ is the sink particle radius, which we set to 2.5 grid cell lengths to be consistent with the \citet{1997ApJ...489L.179T} criterion to avoid artificial fragmentation.

On all AMR levels, except the maximum level, we use an AMR criterion based on the local Jeans length to ensure that $\lambda_\mathrm{J}$ is always resolved with at least 16 grid cell lengths, which is just enough to reasonably resolve turbulent flows with sizes of the order of a Jeans length \citep{2011ApJ...731...62F}.

During an accretion step, the mass, centre of mass and linear momentum of the sink particle are updated by directly following the laws of conservation. However, in order to conserve total angular momentum, an intrinsic angular momentum (spin) has to be introduced. The spin of the sink particle records the accreted angular momentum, to satisfy total angular momentum conservation. The spin is used to determine the angular momentum axis of the sink particle \citep[see][]{2014ApJ...790..128F}, and we are using the same information here to define the sub-resolution orientation of the accretion disc around the sink particle.

We use a multi-grid Poisson solver to compute the self-gravity of the gas \citep{2008ApJS..176..293R}. All gravitational interactions of the sink particles between each other and with the gas are computed by direct summation over all the sink particles and grid cells. A second-order leapfrog integrator is used to advance the sink particles in time.

\subsection{Equation of state (EOS)}
\label{sec:polytropic}
To model the thermodynamics of the gas, we use the method in \cite{2017JPhCS.837a2007F} and employ a polytropic equation of state for the gas pressure $P=P_\mathrm{EOS}$, with
\begin{equation}
    P_\mathrm{{EOS}} = c^2_s\, \rho^{\Gamma}.
\end{equation}
Using the ideal gas EOS, the respective temperature is given by
\begin{equation}
    T_{\mathrm{EOS}} =  \frac{\mu\, m_{\mathrm{H}}}{k_\mathrm{B}\, \rho}\, P_\mathrm{{EOS}} = \frac{\mu\, m_{\mathrm{H}}}{k_\mathrm{B}}\, c^2_s\, \rho^{\Gamma-1}\,.
\end{equation}
Here $c^2_s=(0.2\,\mathrm{km/s})^2$ is the square of the isothermal sound speed for solar-metallicity, molecular gas at $10\,\mathrm{K}$, and $\mu = 2.35$ is the mean molecular weight (in units of hydrogen atom mass $m_{\mathrm{H}}$). The polytropic exponent is defined as
\begin{equation}
    \Gamma =
      \begin{cases}
        1   & \text{for \hspace{7mm} $\rho \le \rho_1 \equiv 2.50 \times 10^{-16}\, \mathrm{g\, cm^{-3}}$,}\\ 
        1.1 & \text{for\, $\rho_1 < \rho \le \rho_2 \equiv 3.84 \times 10^{-13}\, \mathrm{g\, cm^{-3}}$,}\\ 1.4 & \text{for\, $\rho_2 < \rho \le \rho_3 \equiv 3.84 \times 10^{-8}\, \mathrm{g\, cm^{-3}}$,}\\
        1.1 & \text{for\, $\rho_3 < \rho \le \rho_4 \equiv 3.84 \times 10^{-3}\, \mathrm{g\, cm^{-3}}$,}\\
        5/3 & \text{for  \hspace{7mm} $\rho > \rho_4$.}
      \end{cases}
\end{equation}
The value of the polytropic exponent $\Gamma$ varies with the local density of the gas, covering the phases of isothermal contraction, adiabatic heating during the formation of the first and second core and the effects of $\mathrm{H_2}$ dissociation in the second collapse \citep{2000ApJ...531..350M,1969MNRAS.145..271L,1993ApJ...411..274Y,2009ApJ...703..131O}. However, it does not take into account the radiative heating feedback from protostars, which is introduced next.

\section{Stellar heating feedback model} \label{sec:model}

During star cluster formation events, radiation feedback from stars has a significant impact on the immediate environment surrounding the stars, as a result of their high accretion luminosities during the initial phases of formation. For a fully accurate treatment of stellar heating, the radiative transfer equation has to be solved, and this involves tracing rays to every cell around the sink particle. Numerical algorithms that do this are still in their infancy and are currently under development \citep{peters2011radiative,Kuiper_2015,2016NewA...43...49B,2016ApJ...823...28K,2019ApJ...887..108R}. Moreover, in large-scale simulations of star cluster formation, the radiation feedback and associated radiative transfer problem has to be solved for every star and at every timestep of the simulation, making this an extremely hard and computationally expensive problem. Most importantly, in order to obtain the correct radial and angular dependence of the radiation field, one must resolve the individual accretion discs sufficiently well, which is currently still computationally impossible if one wants to follow the entire formation of a star cluster. It has been done for single stars or binaries, but not for entire star clusters, the latter of which is required to obtain a statistically meaningful sample of the IMF.

In order to overcome these problems, we have developed a sub-resolution model that approximates the direct stellar heating from the protostars by assuming a density distribution in the accretion disc surrounding each star. Our new heating feedback model takes into account the radial and angular distances from the stars and models the resulting shielding of the stellar radiation due to the extinction by dust particles.

\subsection{Geometry of the accretion disc}
\begin{figure}
    \centering
    \includegraphics[width=\columnwidth]{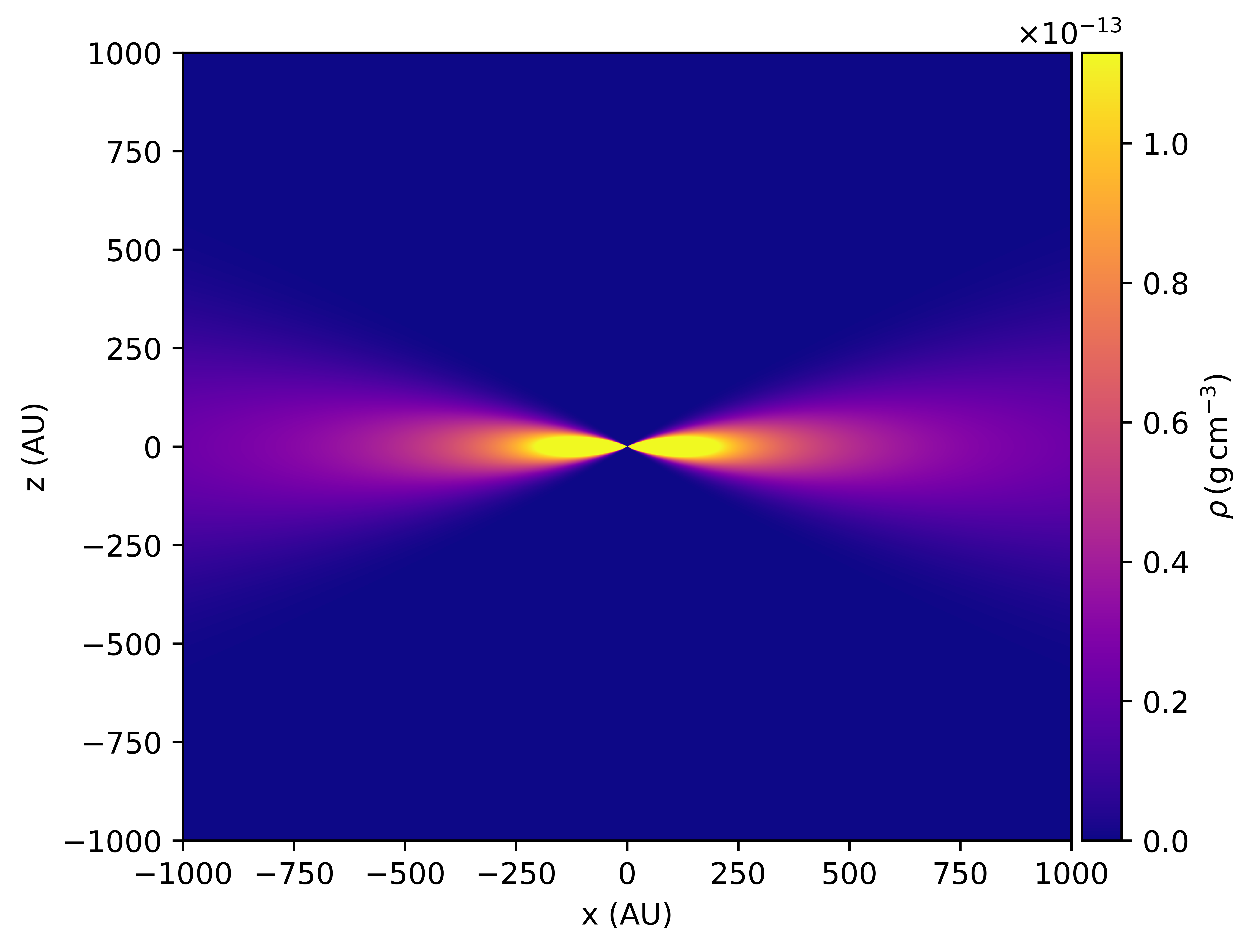}
    \caption{Dust density structure modelled via Eqs.~(\ref{eq:disc1})--(\ref{eq:disc4}), perpendicular to the disc midplane \citep[based on the works of][]{2004A&A...417..793P,2016NewA...43...49B}.}
    \label{fig:dust_dens}
\end{figure}

To approximate the gas and dust density distribution around a young star, we follow the accretion disc models used in \citet{2004A&A...417..793P} and \citet{2016NewA...43...49B}. The dust density distribution is given by
\begin{align}
    \rho(a,z)& = \rho_0\, f_1(a)\, f_2(a,z),\;\mathrm{with} \label{eq:disc1} \\
    f_1(a) & = (a/a_\mathrm{d})^{-1}, \label{eq:disc2} \\
    f_2(a,z) & = \exp\left(-\frac{\pi}{4}\left(z/h(a)\right)^2\right), \label{eq:disc3} \\
    h(a) & = z_\mathrm{d}\,(a/a_\mathrm{d}). \label{eq:disc4}
\end{align}
Here $a=\sqrt{x^2 + y^2}$ is the radial distance in the disc midplane, $z$ is the height above the disc and $\rho_0$ is the density in the disc midplane ($z=0$) at $a_\mathrm{d}=500\,\mathrm{AU}$. A measure for the disc scale height is given by $z_\mathrm{d}=0.25\,a_\mathrm{d}$ \citep{2004A&A...417..793P} and $\rho_0$ is adjusted to resemble the density distribution of an active accretion disc in the protostellar phase. Moreover, inside the inner radius $a_{\mathrm{in}}=1\,$AU, the density is assumed to be zero, i.e., $a_{\mathrm{in}}$ approximates the dust sublimation radius. Fig.~\ref{fig:dust_dens} shows this dust density distribution, perpendicular to the disc midplane. We note that we ignored the slight flare in the shape of the accretion disc when we defined $h(a)\propto a$ in Eq.~(\ref{eq:disc4}).

\subsection{Calculating the total optical depth and the stellar heating temperature}
\label{sec:model_temp}
Following the works of \citet{2002A&A...395..853D} and \citet{2016NewA...43...49B}, the radiative flux density $F_{\star}(r,\theta)$ at every point around the stellar source is given by
\begin{equation}
   F_{\star}(r,\theta) = \frac{L_{\star}}{4\pi r^2}\, \exp\left(-\tau (r,\theta)\right),
\end{equation}
where $r$ is the radial distance and $\theta$ is the angle measured from the angular momentum axis of the sink particle (disc + star, see \S\ref{sec:sink}). $L_{\star}$ is the star's luminosity, which we calculate by using the (proto)stellar evolution model developed by \cite{2009ApJ...703..131O}. The total optical depth
 \begin{equation}
     \tau = \int \kappa\, \rho\, \mathrm{d}r, 
 \end{equation}
where $\kappa$ is the grey opacity and $\rho$ is the dust density distribution.

We calculate the total optical depth by substituting $z=r\,\cos\theta$, $a=r\,\sin\theta$, and  $z_\mathrm{d}=0.25\,a_\mathrm{d}$ in Eqs.~(\ref{eq:disc1})--(\ref{eq:disc4}), to obtain
 \begin{align}
     \tau &= \frac{\kappa\, a_\mathrm{d}\, \rho_0\,    \exp(-4\pi \cot^2\theta)}{\sin\theta} \int_{a_\mathrm{{in}}}^{r} \frac{1}{r}\, dr \nonumber \\
      &= \frac{\kappa\, a_\mathrm{d}\, \rho_0\,    \exp(-4\pi \cot^2\theta)\, \ln(r)}{\sin\theta} \nonumber \\
      &= \frac{\tau_0\, \exp(-4\pi \cot^2\theta)\, \ln(r)}{\sin\theta}\,,
 \end{align}
where $\tau_0=\kappa\, a_\mathrm{d}\, \rho_0\,$ is a constant defined by the opacity and the geometry of the disc. Thus, convenient values of $\tau_0$ can be used to approximate accretion discs in different evolutionary phases. The influence of stellar heating on the surrounding environment is appreciable only during the early stages of the main accretion phase, i.e., when the luminosity is high. Therefore we take $\rho_0=10^{-14}\, \mathrm{g\, cm^{-3}}$, which may be suitable for class 0 or class \MakeUppercase{\romannumeral 1}\, young stellar objects (YSOs).

Finally, we have to compute the heating. The amount of energy absorbed per unit time by a dust particle is
\begin{equation}
    Q (r, \theta) = \chi\, F_{\star}(r,\theta), \label{eq:Qheat}
\end{equation}  
where $\chi = \rho (r, \theta)\, \kappa$ is the absorption coefficient. The dust grains will acquire an equilibrium temperature such that they emit the same amount of energy they absorb. Neglecting the reprocessed radiation field, we find
\begin{equation}
    \frac{\sigma_\mathrm{{SB}}}{\pi}\, \chi\, T_{\mathrm{heat}}^4 = \frac{Q}{4\pi},
\end{equation}
where $\sigma_\mathrm{{SB}}$ is the Stefan-Boltzmann constant and $T_{\mathrm{heat}}$ is the temperature due to stellar heating. Thus, the final gas temperature is given by the sum of the energies \citep[temperature to the 4th power; see][]{2016MNRAS.458..673G,2018AAS...23111403G,2017JPhCS.837a2007F} due to the EOS and the stellar heating,
\begin{equation}
    T = {(T^4_{\mathrm{EOS}} + T^4_{\mathrm{heat}})}^{1/4}. \label{eq:Theat}
\end{equation}
We can also express this in terms of the total gas pressure,
\begin{equation}
    P = \left[P^4_{\mathrm{EOS}} + {\left(\frac{k_B\, \rho}{\mu\, m_\mathrm{H}}\right)}^4\, T^4_{\mathrm{heat}}\right]^{1/4}, \label{eq:Pheat}
\end{equation}
which is applied in the MHD momentum equation, Eq.~(\ref{eq:mhd2}).

\section{Initial conditions and simulation parameters}
\label{sec:parameters}

The simulations are carried out in a three-dimensional triple-periodic computational box with side length $L=2\mathrm{pc}$. The maximum refinement level gives a maximum effective grid resolution of $N_{\mathrm{eff,\, res}}^3=2048^3$ cells or a minimum cell size of $\Delta x_{\mathrm{cell}}=200\,\mathrm{AU}$. The total cloud mass is $M=775\, \mathrm{M_{\odot}}$, with an initial uniform gas density $\rho = 6.56 \times 10^{-21}\, \mathrm{g\, cm^{-3}}$ and a mean freefall time of $t_{\mathrm{ff}}= 0.82\,$Myr. The turbulence driving creates local compressions or shocked regions, leading to the fragmentation of the cloud and the formation of filaments where dense cores are formed. These cores are the sites of star formation \citep{2013A&A...551C...1S,2014prpl.conf...27A}. The velocity dispersion $\sigma_v=1.0\, \mathrm{km\, s^{-1}}$ and the initially isothermal sound speed $c_s=0.2\, \mathrm{km\, s^{-1}}$ sets the amplitude of the driving of the turbulence with a steady-state sonic Mach number $\mathcal{M}=\sigma_v/c_s=5.0$. The magnetic field is uniform initially with $B= 10^{-5}\, \mathrm{G}$ along the z-axis of the computational domain, which is also modified by the turbulence, approximating the structure of magnetic fields in real molecular clouds \citep{2016JPlPh..82f5301F}. The initial virial parameter $\alpha_\mathrm{{vir}}=2E_\mathrm{{kin}}/E_\mathrm{{grav}}=0.5$ is in the range of observed values \citep{1992A&A...257..715F,2013ApJ...779..185K,2015ApJ...809..154H}. The gas in the box is initially stirred in the absence of self-gravity. A fully-developed turbulent state is reached after two turbulent crossing times, $2 t_\mathrm{turb}=L/(\mathcal{M}c_s) = 2\,\mathrm{Myr}$, which is when we activate self-gravity and sink particles. We study the time evolution of different dynamical quantities and the IMF from this point in time, which we define as $t=0$, i.e., when self-gravity is turned on. This procedure is similar to that used in previous works \citep[e.g.,][]{2012ApJ...761..156F,2016ApJ...822...11P}.

We compare three models with different realisations of gas heating: 1) polytropic, 2) spherical, and 3) polar heating (see Table~\ref{tab:sims}). The polytropic simulation does not include the radiative feedback from stars and heating occurs only due to gas compression (see \S\ref{sec:polytropic}). The spherical model approximates the stellar heating by assuming spherical symmetry and a homogeneous distribution of the gas. According to this model, the flux density $S_{\star}$ at any point around the protostar is given by \citet{2017JPhCS.837a2007F},
\begin{equation}
   S_{\star}=\frac{L_{\star}}{4\pi r^2}.
\end{equation}
In this model, the temperature due to the heating by a star of given luminosity is calculated as in Eqs.~(\ref{eq:Qheat})--(\ref{eq:Pheat}), but with $F_{\star}$ only a function of radial distance. In contrast, the polar model takes into account the structure and orientation of the accretion disc around each protostar and the extinction by the dust grains in the disc.

\begin{table*}
	\centering
	\caption{Key simulation parameters and results}
	\label{tab:sims}
	\begin{tabular}{lccccccc} 
	    \hline
		\hline
		 Heating model & $N_{\mathrm{eff,\,res}}^3$ & $\Delta x_{\mathrm{cell}}$ [AU] & $r_{\mathrm{sink}}$ [AU] & $\rho_{\mathrm{sink}}$ [$\mathrm{g\ cm^{-3}}$] & $N_{\mathrm{sims}}$& $N_{\mathrm{sinks}}$ & Average $M_{\mathrm{sink}}$ [$\mathrm{M_{\odot}}$] \\
        (1) & (2) & (3) & (4) & (5) & (6) & (7) & (8) \\
		\hline
		\hline
		1. Polytropic heating & $2048^3$ & 200 & 500 & $8.3 \times 10^{-17}$ & 10 & 305 & 2.54\\
		2. Spherical heating  & $2048^3$ & 200 & 500 & $8.3 \times 10^{-17}$ & 10 & 206 & 3.69\\
		3. Polar heating      & $2048^3$ & 200 & 500 & $8.3 \times 10^{-17}$ & 10 & 271 & 2.87\\
		\hline
	\end{tabular}
\end{table*}

\section{Results}
\label{sec:results}

The use of sub-resolution models in simulations to approximate the radiative heating by stars is primarily aimed at facilitating parameter studies, particularly when numerous simulations have to be performed for better statistics. Due to the chaotic nature of the turbulence, the simulation results may vary when repeated with the same physical setup, but different random seeds of turbulence. Thus, in order to obtain statistically meaningful results, one must average over many different realisations of the same parameter set. Using our new sub-resolution model for stellar feedback, we can carry out many cloud simulations and analyse the statistical quantities from the aggregate data. Here we run and analyse 10 simulations for each of the three heating models listed in Tab.~\ref{tab:sims}

\subsection{Column density and temperature structure}
\begin{figure*}
    \centering
    \includegraphics[width=\textwidth]{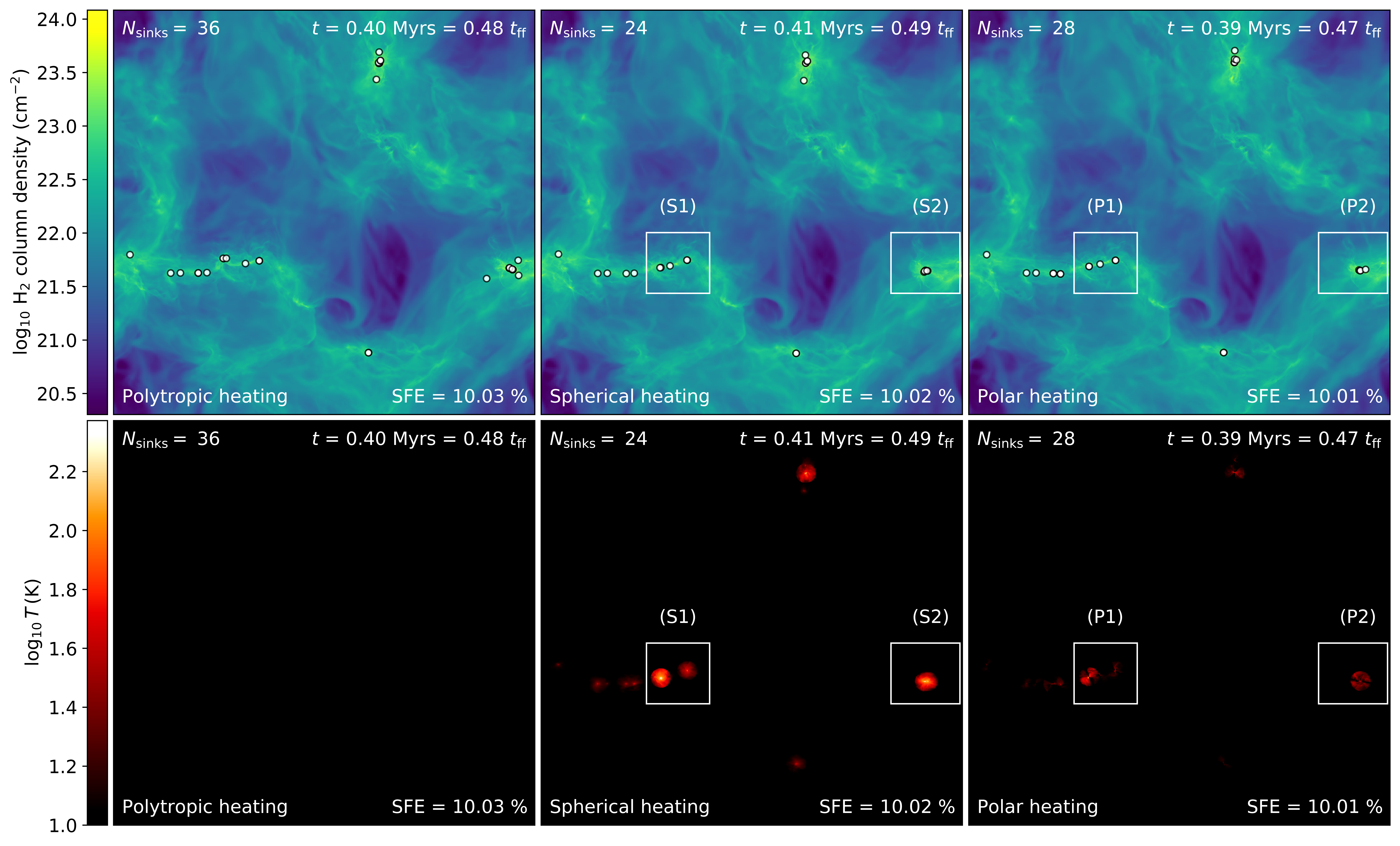}
    \caption{The top panels show the column density along the z-axis and the ones on the bottom panels show the temperature maps at SFE = 10\%. Each column corresponds to simulations with a polytropic EOS without radiative feedback (left), polytropic EOS combined with the spherical stellar heating feedback (middle), and polytropic EOS combined with the polar heating feedback (right). The white circles in the top panels represent the positions of sink particles. Almost no heating occurs for the purely polytropic EOS, while the spherical and polar heating models produce local temperatures around each protostar of up to several hundred K. Zoom-in regions are marked with squares in the spherical and polar runs labelled S1, S2 and P1, P2, respectively, and shown in Fig.~\ref{fig:zoom}.}
    \label{fig:maps}
\end{figure*}

Fig.~\ref{fig:maps} shows one particular realisation of the spatial distribution of the column density (the gas number density integrated along the line-of-sight) and the gas temperature for simulations with the heating models 1 (polytropic), 2 (spherical), and 3 (polar) from Tab.~\ref{tab:sims} at the time when the star formation efficiency SFE = $10\%$ is reached (i.e., a fraction of $10\%$ of the total cloud mass has formed stars). The polytropic, spherical and polar heating simulations form 36, 24, and 28 sink particles, respectively. There is almost no heating in the polytropic simulation because feedback from the stars is completely ignored in that model. In contrast, local heating around newly formed protostars up to several hundred K occurs in both the spherical and polar heating runs. However, the polar heating model heats less and primarily in the directions along the rotation of the sink particles as intended.

The reduced heating in the polar simulation results in more fragmentation in the cloud cores compared to the spherical heating model. This leads to a higher number of stars formed in the polar heating simulation. However, the number of formed sink particles is highest in the polytropic simulation, with 36 sinks formed. In the polytropic model, heating occurs only due to the thermal evolution of the gas, and the stellar heating feedback is absent. As a result, the temperature is almost uniform at $\sim 10$ K. The heating by the stars is spherically symmetric in the spherical simulations, and high temperatures are attained close to the sink particles. We calculate $T_{\mathrm{heat}}$ only up to distances of $10^4$ AU from the sink particles as is the range in which the stellar heating is most important for low- and intermediate-mass stars \citep{2009ApJ...703..131O}. In the case of the polar heating model, heating is restricted to the region surrounding the polar axis of the sinks and there is hardly any increase in the temperature in the regions where accretion discs would form. We note our simulations do not resolve the discs, but instead a geometry is assumed based on the radial distance from the sinks and the angular distance from the angular momentum axis of the sinks; see \S\ref{sec:model}).

\begin{figure*}
    \centering
    \includegraphics[width=\textwidth]{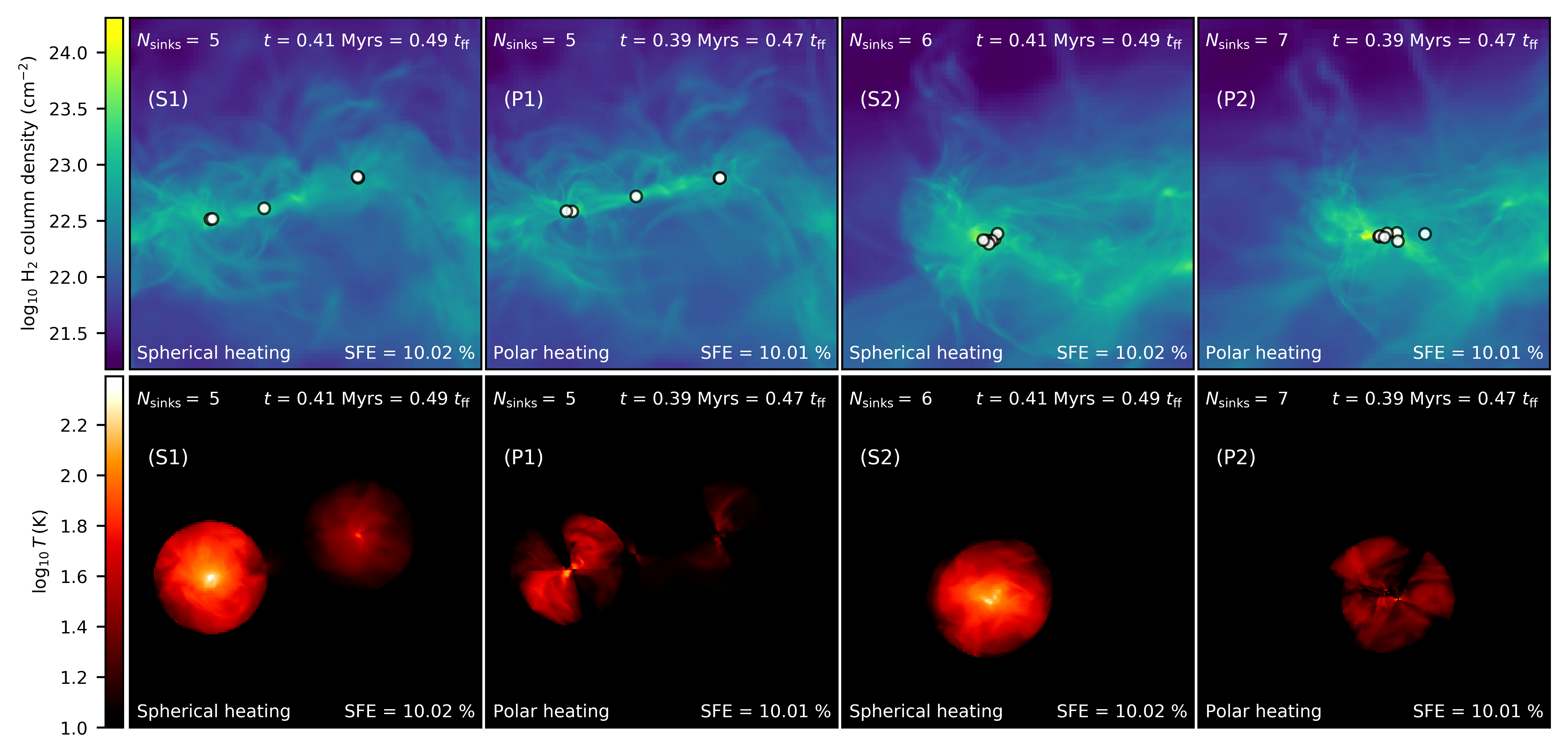}
    \caption{The first and second columns represent the zoomed-in images of the region within S1 and P1 of the spherical and polar simulations, respectively, in Fig.~\ref{fig:maps}. The third and fourth columns represent the same for the region within S2 and P2 in Fig.~\ref{fig:maps}.}
    \label{fig:zoom}
\end{figure*}

 Fig.~\ref{fig:zoom} presents zoomed-in images of the regions within the square outlines in Fig.~\ref{fig:maps}. In the region interior to S1 and P1 (columns 1 and 2 in Fig.~\ref{fig:zoom}), the stars are positioned along a filament and somewhat spaced from one another, with the exception of pairs (binary stars). Although there are slight variations in the density distribution, both the spherical and polar simulations concur in terms of the number of sinks formed. One can clearly see the difference in the temperature structure produced by the spherical and polar heating models. The region within S2 and P2 from Fig.~\ref{fig:maps} (columns 3 and 4 in Fig.~\ref{fig:zoom}) have a clustered arrangement of stars, where the stars are in close proximity to each other. There are more pronounced differences in the spatial distribution and number of the sink particles formed between the two models. In the spherical heating model, the heating regions of different sink particles overlap and the surrounding gas is heated to higher temperatures, such that no new stars can form there. Such a situation, which occurs in simulations with high star formation rates, has been mentioned in \cite{2011ApJ...740...74K}. In contrast, in case of polar heating, due to the difference in the orientation of the accretion discs, the overlap of the heating zones and therefore the temperature of the surrounding gas are significantly reduced. This allows for the formation of an additional sink particle (see on the right, next to the cluster of stars in column 4 of Fig.~\ref{fig:zoom}). These results demonstrate the importance of accurately modeling the stellar heating in simulations of star cluster formation.
 
 \subsection{Evolution of dynamical quantities}
\begin{figure*}
    \centering
    \includegraphics[width=\textwidth]{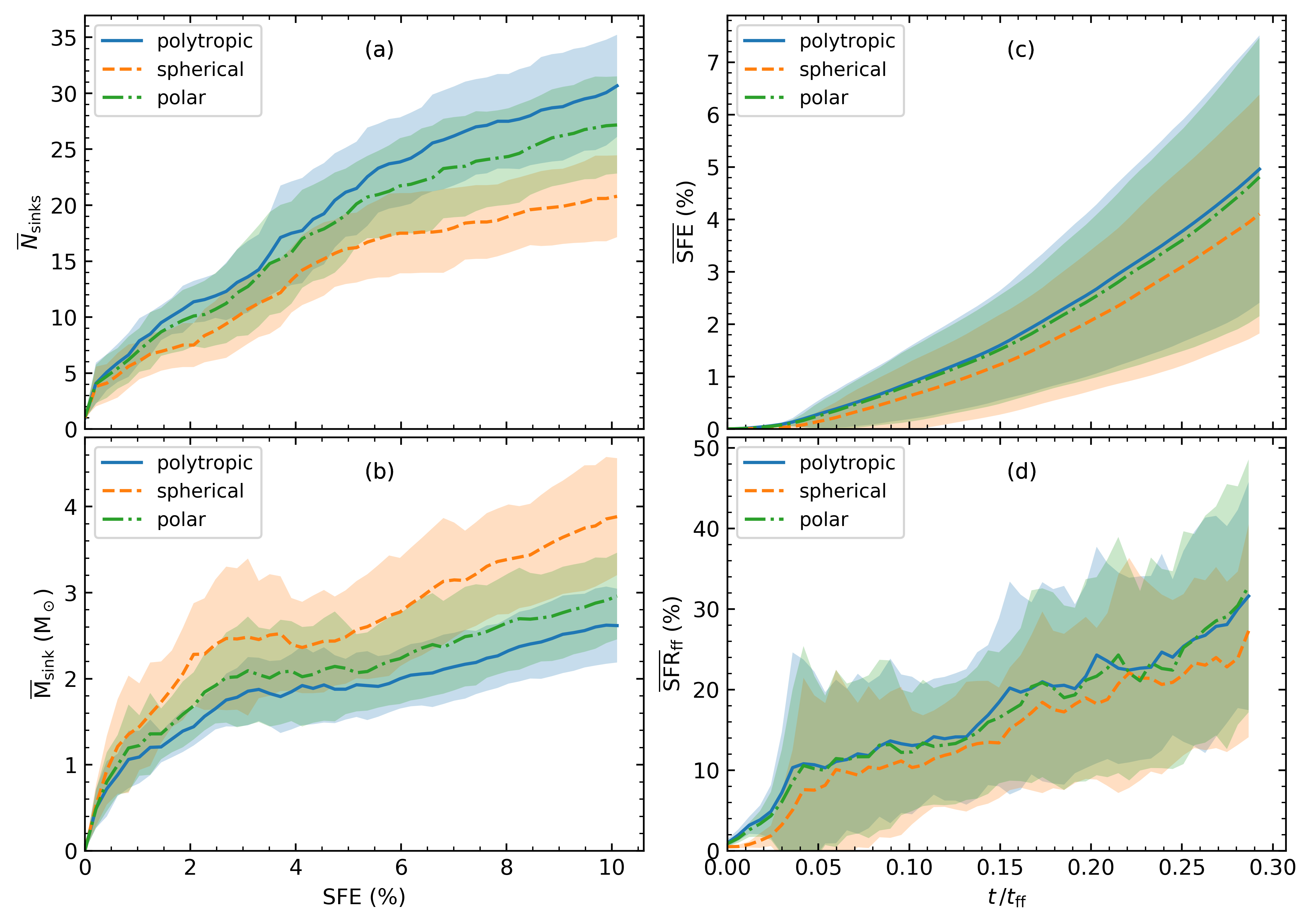}
    \caption{The left panels show (a) the average number of sink particles formed and (b) the average stellar mass as a function of the star formation efficiency (SFE in \%). The right panels indicate the time evolution of (c) average SFE and (d) average star formation rate per freefall time ($\mathrm{SFR_{ff}}$ in \%). All quantities shown here represent the average values obtained from $10$ simulations with different turbulence seeds, and the coloured bands correspond to the standard deviation over the sample of the $10$ simulations, for each of the three heating models.}
    \label{fig:Plots}
\end{figure*}

 For each of the three heating models, we run 10 simulations with different random seeds of the turbulence (see Tab.~\ref{tab:sims}), and analyse the evolution of the statistical values of dynamical quantities. Fig.~\ref{fig:Plots} shows the evolution of the number of sink particles formed, the average stellar mass, the star formation efficiency and the star formation rate per freefall time ($\mathrm{SFR_{ff}}$). The latter is the percentage of the total cloud mass that has formed stars per unit time, where time is expressed in units of the freefall time at the cloud mean density. For the number of sink particles as a function of SFE, all three heating models follow almost the same trend at lower SFEs, but start to deviate as the value of SFE increases. The reason for this is that as the number of stars formed increases, the impact of stellar heating feedback on the parent cloud becomes more pronounced. The bottom-left panel displays the change in the average sink particle mass with SFE. The polytropic and polar models follow similar curves, but the average sink particle mass in the spherical model tends to higher values with increasing SFE compared to the polytropic and polar models. Finally, both the SFE and $\mathrm{SFR_{ff}}$ do not significantly depend on the choice of heating model, as evident from the right-hand panels. Thus, the main effect of stellar heating is not on the accretion rate, but on the fragmentation of the gas.

\subsection{Initial mass function}
\begin{figure}
    \centering
    \includegraphics[width=\columnwidth]{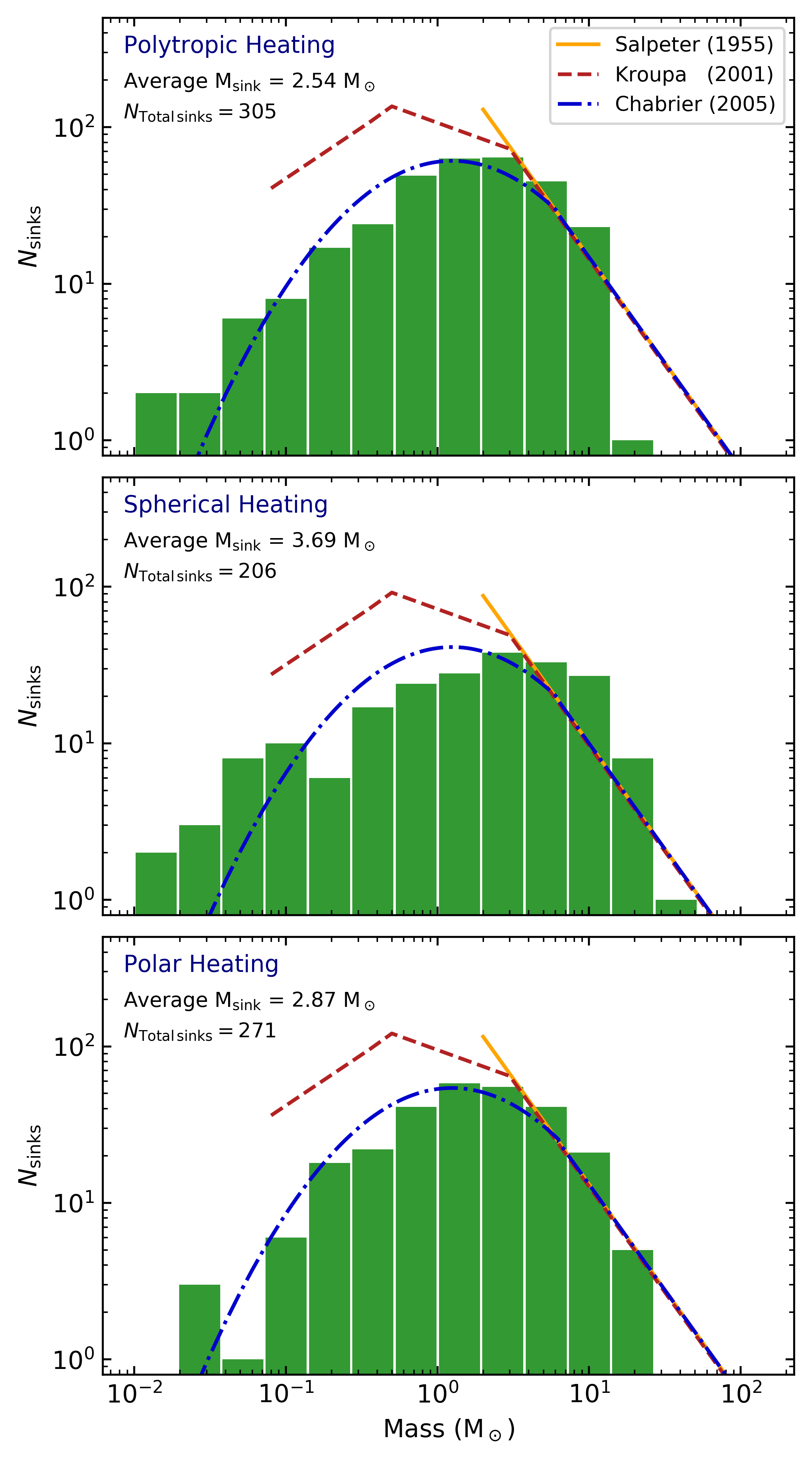}
    \caption{Initial mass function (IMF) of sink particles (stars) formed in the simulations with the three different heating models (polytropic: top; spherical: middle; polar: bottom). Each distribution contains data from $10$ simulations with different turbulence realisations. The observed IMF models by \citet{1955ApJ...121..161S} (solid), \citet{2001MNRAS.322..231K} (dashed), \citet{2005ASSL..327...41C} (dashed-dotted) have been shifted by a factor of $\sim 6$ to match the peak of the mass distribution from the simulation for better comparison. The main reason for this mismatch in characteristic mass is that the present simulations do not include the effects of outflows and jets, which significantly reduce the characteristic mass \citep{2007A&A...462L..17A,2014ApJ...790..128F,2014prpl.conf..243K}. We also note that the \citet{2001MNRAS.322..231K} model corresponds to the canonical IMF (all stars counted); however, binaries are not resolved in our simulations \citep[see][]{2013pss5.book..115K}.}
    \label{fig:imf}
\end{figure}

 Our work differs from other studies of the IMF \citep[e.g.,][]{2009MNRAS.392.1363B,2014MNRAS.439.3420M,2018MNRAS.476..771C} in terms of the availability of a statistically representative sample obtained from many turbulent realisations of the same cloud. Fig.~\ref{fig:imf} shows the histograms of the initial distribution of stellar masses in $10$ simulations of each of the three heating models. The inclusion of the heating feedback in the simulations (spherical and polar) resulted in the formation of a higher number of stars in the high-mass end compared to the polytropic heating model. The additional heating from stellar feedback suppresses the fragmentation of the cloud core, allowing fewer stars to accrete more gas. Because of this, the spherical and polar heating model seem to better reproduce the Salpeter slope at the high-mass end of the distributions, while the purely polytropic EOS produces too few high-mass stars. The polar heating model matches the observed high-mass tail the best of all the three heating models. Furthermore, we see that the characteristic mass in the spherical heating model is higher than that in the other models. This may be a consequence of the overheating problem proposed by \cite{2011ApJ...740...74K}, leading to a top-heavy IMF. As a result of the high star formation rates in our simulations (see Fig.~\ref{fig:Plots}), the spherically-symmetric heating regions overlap and over-suppress the expected fragmentation of the surrounding gas. In the polar heating model, although many stars form close to each other, the overlap is significantly reduced compared to the spherical heating model, due to the asymmetry of the heating introduced by the disc sub-resolution model (see \S\ref{sec:model}).
 
 Although the mass distribution of stars in the polytropic runs appear to reproduce the observed shape in the low-mass domain, in the resolution-study performed by \cite{2017JPhCS.837a2007F}, it was shown that high-resolution runs with the polytropic heating model produced an exceedingly high number of low-mass stars and brown dwarfs. The absence of radiative heating results in immoderate fragmentation on small-scales close to stars \citep{2016MNRAS.458..673G,2018AAS...23111403G}, leading to unreasonable production of low-mass objects \citep{2009MNRAS.392.1363B,2009ApJ...703..131O}. The polar heating model, comparatively, achieves the best convergence on the overall shape of the IMF.
 
\section{Limitations}
\label{sec:discussions}

\subsection{Jet and outflow feedback}
In our simulations, the module for producing the effects of mechanical outflows and jets from protostars \citep{2014ApJ...790..128F} was not included, since the preliminary focus of this study is on establishing the importance of stellar radiation feedback in controlling the IMF. This explains why the peak mass in all our simulations is higher ($ >1\, \mathrm{M_\odot})$ than in the observed IMF. In the comparison studies of \cite{2018MNRAS.476..771C}, a mass distribution with a similar peak mass $(\sim1\,\mathrm{M}_\odot)$ was observed for a simulation model without jet/outflow feedback, while the same model with jet/outflow feedback produced a characteristic mass comparable to the observed IMF peak. The ejection of matter through the bipolar outflows limits accretion and results in additional fragmentation \citep{2014ApJ...790..128F}. This in turn leads to a reduction in the stellar masses by a about a factor of $\sim 3$ \citep{Li_2010,2014ApJ...790..128F}. We note that although the mass distribution produced by the polar heating model matches the overall shape of the observed IMF reasonably well, it is possible that the shape may be altered, particularly at the low-mass end, with the inclusion of mechanical outflows.

\subsection{Numerical resolution}
The resolution in our simulations was not high enough to resolve protostellar discs. Higher-resolution runs may permit possible fragmentation on the disc scales, and the formation of a higher number of stars. Further, it may be relevant to study the combined effects of outflow and radiative feedback. The ejected matter from stars or young stellar objects sweeps away the surrounding envelope of gas, forming cavities. Therefore, the influence of radiative feedback may be modulated by the jet/outflow feedback \citep{2019FrASS...6....7K}.

\section{Conclusions}
\label{sec:conclude}
We have implemented a simple direct stellar heating module in the FLASH MHD code. The module approximates the effects of energy transfer in the form of radiation from stellar sources, also considering the loss in intensity of the radiation field due to the absorption by dust grains in the disc surrounding each protostar. The implementation of such sub-resolution models allow us to perform large parameter studies and overcome computational cost and limitations otherwise present in simulations of full radiation transfer. 

We carry out a set of MHD simulations with different models for the evolution of the gas thermodynamics: 1) polytropic (heating only due to gas compression), 2) polytropic plus spherically-symmetric stellar heating, and 3) polytropic plus polar stellar heating, which considers the extinction of stellar radiation by the dust particles in the protostellar accretion discs. We compare the spatial distribution of formed stars and their initial mass functions in 10 simulations for each of the three different models. We demonstrate that stellar radiative feedback has a prominent effect on the number of stars formed in the cluster, and the extent of influence conspicuously depends on the degree of overlap of the stellar heating zones. In particular, the density distribution and the number of stars vary between the heating models in regions where there is significant overlap of heating zones, i.e., crowding of stars. The excessive overlap in the spherical model hinders any potential fragmentation which leads to a high characteristic mass, and eventually, a top-heavy IMF. However, in the case of the polar heating model, we take into consideration the existence of optically-thick accretion discs around the young stars, which results in the confinement of heating to the polar directions due to the shielding of radiation by the dust particles in the discs. This significantly reduces the overlap of the heating zones in the polar model and leads to a more realistic shape of the resulting IMF.

It was also observed that the time evolution of the SFE and SFR does not vary significantly between the three heating models, implying that the accretion rate is rather unaffected by stellar heating, at least when only considering low- and intermediate-mass stars.

We find that both the spherical and polar heating models produce more high-mass stars than the polytropic ones due to the reduction in fragmentation. We further show that of the three heating models tested, the polar heating model achieves the best match to the overall shape of the IMF. However, it is not certain from the current studies that the shape would be retained once important additional physics like jets and outflows are included. This will be addressed in follow-up studies, where similar simulations would be performed, but of higher resolution and including outflow feedback. We conclude that in simulations of star cluster formation, accurate modeling of the stellar heating feedback is necessary to obtain physically meaningful results and is fundamental to the understanding of the stellar IMF.

\section*{Acknowledgements}
We thank the anonymous reviewer for their comments, which helped to improve the paper. C.~F.~acknowledges funding provided by the Australian Research Council (Discovery Project DP170100603 and Future Fellowship FT180100495), and the Australia-Germany Joint Research Cooperation Scheme (UA-DAAD). We further acknowledge high-performance computing resources provided by the Leibniz Rechenzentrum and the Gauss Centre for Supercomputing (grants~pr32lo, pr48pi and GCS Large-scale project~10391), the Australian National Computational Infrastructure (grant~ek9) in the framework of the National Computational Merit Allocation Scheme and the ANU Merit Allocation Scheme. The simulation software FLASH was in part developed by the DOE-supported Flash Center for Computational Science at the University of Chicago.

\section*{Data availability}
The data used in this article is available upon request to the authors.




\bibliographystyle{mnras}
\bibliography{Bibliography} 



\bsp	
\label{lastpage}
\end{document}